\documentclass[%
reprint,
superscriptaddress,
mathtools,amsmathm,amssymb,
 aps,
]{revtex4-2}

\usepackage{graphicx}
\usepackage{dcolumn}
\usepackage{bm}

\usepackage[dvipsnames]{xcolor}
\usepackage{sidecap}
\usepackage{float}
\usepackage{amsmath}

\begin{document}

\preprint{APS/123-QED}

\title{An Electromagnetic Approach to Cavity Spintronics}
\author{Rair Mac\^edo*}
\author{Rory C. Holland}
\author{Paul G. Baity}
\author{Luke J. McLellan}
\affiliation{James Watt School of Engineering, Electronics \& Nanoscale Engineering Division, University of Glasgow, Glasgow G12 8QQ, United Kingdom}
\email{Rair.Macedo@glasgow.ac.uk}

\author{Karen L. Livesey}
\affiliation{School of Mathematical and Physical Sciences, The University of Newcastle, Callaghan NSW 2308, Australia
}
\affiliation{Center for Magnetism and Magnetic Materials, Department of Physics and Energy Science, University of Colorado Colorado Springs, Colorado Springs, Colorado 80918, USA}

\author{Robert L. Stamps}
\affiliation{Department of Physics and Astronomy, University of Manitoba, Winnipeg, Manitoba, MB R3T 2N2, Canada
}
\author{Martin P. Weides}
\affiliation{James Watt School of Engineering, Electronics \& Nanoscale Engineering Division, University of Glasgow, Glasgow G12 8QQ, United Kingdom}

\author{Dmytro A. Bozhko}
\affiliation{Center for Magnetism and Magnetic Materials, Department of Physics and Energy Science, University of Colorado Colorado Springs, Colorado Springs, Colorado 80918, USA}
\affiliation{James Watt School of Engineering, Electronics \& Nanoscale Engineering Division, University of Glasgow, Glasgow G12 8QQ, United Kingdom}

\date{\today}

\begin{abstract}
The fields of cavity quantum electrodynamics and magnetism have recently merged into \textit{`cavity spintronics'}, investigating a quasiparticle that emerges from the strong coupling between standing electromagnetic waves confined in a microwave cavity resonator and the quanta of spin waves, magnons.
This phenomenon is now expected to be employed in a variety of devices for applications ranging from quantum communication to dark matter detection.
To be successful, most of these applications require a vast control of the coupling strength, resulting in intensive efforts to understanding coupling by a variety of different approaches.
Here, the electromagnetic properties of both resonator and magnetic samples are investigated to provide a comprehensive understanding of the coupling between these two systems.
Because the coupling is a consequence of the excitation vector fields, which directly interact with magnetisation dynamics, a highly-accurate electromagnetic perturbation theory is employed which allows for predicting the resonant hybrid mode frequencies for any field configuration within the cavity resonator.
The coupling is shown to be strongly dependent not only on the excitation vector fields and sample's magnetic properties but also on the sample's shape. 
These findings are illustrated by applying the theoretical framework to two distinct experiments: a magnetic sphere placed in a three-dimensional resonator, and a rectangular, magnetic prism placed on a two-dimensional resonator.
The theory provides comprehensive understanding of the overall behaviour of strongly coupled systems and it can be easily modified for a variety of other systems.
\end{abstract}

\maketitle


\section{Introduction}
$ $

The concept of using electromagnetic waves at millimetre wavelengths trapped within resonators to probe quantum properties of matter is no stranger to us. 
In fact, it dates back to the 1940’s when Purcell and colleagues published an abstract which was later presented at the 1946 Spring Meeting of the American Physical Society \cite{purcell46}.
In that work, they showed that the transitions between energy levels, which correspond to different orientations of the nuclear spin in the presence of a static applied magnetic field, can couple to a resonant circuit. 
This coupling could then be measured through changes in the quality factor of the system. 
Their work was the steppingstone to the field of cavity quantum electrodynamics \cite{walther06}.
Interestingly enough, in that same year Griffiths also used standing waves in a microwave resonator to measure the effective high-frequency permeability of ferromagnets \cite{griffiths46} which then led to Kittel's theory of ferromagnetic resonances \cite{kittel48}.
More recently, these two -- once distinct -- lines of research have come together in a newly designated area of research known as \textit{cavity spintronics} which is concerned with studying `cavity magnon-polaritons' \cite{goryachev14,zhang14,zhang15b}.   
These are hybrid light–matter quasiparticles originating from the strong coupling between magnons (the quanta of spin waves) and electromagnetic waves bound inside a microwave cavity resonator \cite{zhang14}.
One of the most fascinating aspects of these hybrid cavity-magnon systems is the potential to combine light and magnetism; and by doing so it should be possible to combine quantum information with spintronics \cite{tabuchi15,lachance_quirion19}. 
In addition, this emergent phenomenon can also be used to engineer devices including, gradient memory devices \cite{zhang15}, ferromagnetic haloscopes for axion detection \cite{crescini18,flower19,crescini20}, and radiofrequency-to-optical transducers \cite{hisatomi16}.

In order to fully exploit cavity-magnon hybrid quasiparticles for applications, a deep understanding of the coupling strength is required. 
The coupling strength determines the degree of coherent information exchange, and thus, plays a crucial role when constructing any devices employing cavity spintronics.
As an example of recent efforts into fully understanding cavity magnon-polariton coupling, we can quote Zhang and colleagues' findings\cite{zhang17} on the observation of exceptional points (where the two-level system's eigenfrequencies coalesce) in a cavity magnon–polariton system upon tuning the magnon–photon coupling strength.
In addition, the optimisation of the coupling conditions has been shown to be a vital aspect of obtaining non-Markovian dynamics in a multi magnet-cavity hybrid system employed as a coherent, long-lifetime, broadband and multimode gradient memory with a 100-ns storage \cite{zhang15}.
Mechanisms to control the coupling strength have so far included changing the position of the sample within the resonator \cite{harder18}, voltage induced control \cite{kaur16}, as well as varying the temperature of the system \cite{flaig17}.
More recently, a two-port cavity approach has been implemented using two-\cite{bhoi19,zhang19} and three-dimensional\cite{boventer19a,boventer19b} systems as a way to achieve level attraction as well as coherent manipulation of energy exchange in the time domain \cite{wolz19}.
These are only a few examples of the intensifying interests to fully understand and manipulate the coupling behaviour in hybrid cavity spintronic systems.

However, up to now most works have neglected how the excitation vector fields within the resonator can modify the coupling of the hybrid modes and, more importantly, how these fields directly interact with magnetisation dynamics.
This includes the direction and profile of the cavity fields.
A few different models have been used to describe the magnet-cavity system, one of which is the harmonic coupling model. This treats the magnet and cavity as two coupled harmonic oscillators (microscopically \cite{goryachev14} or macroscopically \cite{harder16,proskurin19}). 
Another is the dynamic phase correlation model which looks at impedance changes due to charge motion generated by spin precession inside the cavity -- thus relating the system to Amp\'ere’s and Faraday’s laws \cite{cao19,harder16}.
While these models have captured much of the nature of hybrid cavity-spin systems, they still do not consider the full effect of complex driving fields on the spin dynamics. In addition, they also require the introduction of various experimentally-extracted parameters.

Here, we demonstrate experimentally that by modifying the position of the sample inside a resonator as well as changing the sample's shape, it is possible to drastically change the coupling strength.
We explain the results with an elegant theory based on fundamental magnetic torque equation combined with Maxwell's equations. 
The theory is used for predicting the hybrid magnon-polariton frequencies and it shows remarkable agreement with experimental data without the use of fitting parameters or phenomenological terms such as spin density. 
To demonstrate that the theoretical method is generally applicable to any magnet-cavity system we use two illustrative cases:
we start with a microwave cavity resonator where linearly polarised excitation is obtained, and place a magnetic Yttrium Iron Garnet (Y$_3$Fe$_5$O$_{12}$ or simply YIG) \cite{cherepanov93,serga10} sphere inside. 
We then change the position of the sphere to exemplify how the coupling strength can be drastically modified with small changes in the microwave field profile at the sample position. 
Further, we investigate similar behaviour in a different cavity resonator -- namely a two-dimensional wave guide resonator.
Using a perturbation method, we provide a theoretical framework to describe the behaviour of cavity spintronic systems based on self-consistent electromagnetic theories. 
This allows for an accurate verification of our experimental findings using analytical expressions for the field profile inside the cavity and accounting for its coupling with specific magnetic permeability tensor components. 
This tensor is obtained from the magnetic torque equation (ie. the Landau Lifshitz equation) and can be used to treat magnets of various types and shapes. 
Hence, our theoretical framework is very general and can be tailored to fit a variety of different hybrid systems. 
Finally, we expect that by being able to fully understand the behaviour of these systems, we open up new avenues for exchange and manipulation of information through cavity spintronic devices; in both classical and quantum regimes.

\section{Theoretical Framework}
$ $

Before listing our main findings, it will be necessary to revisit two well-known concepts in magnetism and microwave engineering: the response of  magnetisation to an oscillating magnetic field, characterised through a dynamic susceptibility; and electromagnetic perturbation theory in a microwave resonator. These are essential for a faithful theoretical description of cavity-magnon hybridisation. 

\subsection{Magnetic response through a dynamic susceptibility}\label{sec:LLG}
$ $

Let us start by looking at ferromagnetic resonances. 
This, in general, happens when a steady magnetic field, $\mathbf{H_{ext}}$, is applied to a spin system wherein the total magnetic moment, $\mathbf{M}$, will coherently precess about its equilibrium orientation. 
Resonance will occur when an oscillating magnetic field is applied with frequency equal to that of the natural Larmor frequency of the magnet.
The behaviour can be semi-classically described by the equation of motion of magnetisation (the Landau-Lifshitz equation) \cite{gurevichBook}:

\begin{align}
\label{llg}
\frac{\partial\mathbf{M}}{\partial t} = -\gamma\mu_0(\mathbf{M}\times \mathbf{H_{0}}).
\end{align}
Here, the magnetisation is given by $\mathbf{M} = \hat{\mathbf{z}}M_s + \mathbf{m}e^{j\omega t}$, with $M_s$ being the saturation magnetisation, $\gamma$ is the gyromagnetic ratio, and $\omega$ is the angular frequency. 
Note that magnetic damping is ignored for now but it will later be taken into account phenomenologically. 
The effective field, $\mathbf{H_{0}}$, acting on $\mathbf{M}$ includes contributions from the various energy terms such as Zeeman, dipole-dipole, exchange and anisotropy. 
Here we consider that it contains terms due to the oscillating field $\mathbf{h}$ and the externally applied magnetic field $\mathbf{H_{0}}$ along the $z$ direction. 
To account for the shape of magnetic samples, we also include contributions from a demagnetising field which can be written as $\mathbf{H_{D}} = -\overleftrightarrow{D}\cdot\mathbf{M}$, where $\overleftrightarrow{D}$ denotes the demagnetising tensor $diag(D_x,D_y,D_z)$ \cite{brown62}.
The effective field can then be written as $\mathbf{H_{eff}} = \mathbf{H_D}+\hat{\mathbf{z}}H_{0}+\mathbf{h}e^{j\omega t}$.

After applying these definitions to Eq.~(\ref{llg}), one arrives at the relation between the oscillating magnetisation, $\mathbf{m}$, and the oscillating magnetic field,  $\mathbf{h}$:

\begin{align}
\label{mhs}
\begin{bmatrix}
m_x \\
m_y \\
\end{bmatrix} =
\underbrace{\begin{bmatrix}
\chi_{xx} & j\chi_{xy} \\
-j\chi_{yx} & \chi_{yy} \\
\end{bmatrix}}_{\overleftrightarrow{\chi}_m(\omega)}
\begin{bmatrix}
h_x \\
h_y \\
\end{bmatrix}.
\end{align}
Here, $\overleftrightarrow{\chi}_m(\omega)$ is the high-frequency magnetic susceptibility which is a second rank tensor. This tensor is often used to describe the electromagnetic response of magnetic materials. It is noteworthy that the nonzero off-diagonal elements are well-known to give rise to various nonreciprocal effects \cite{camley87,macedo19}, which are the basis for a number of important device applications \cite{how05,fuller1987}. 

Before looking at the behaviour of a magnet inside a microwave cavity, some intuition can be gained by exploring Eq.~(\ref{mhs}) under a few different circumstances. 
For this we will look at $\chi_{xx}$, which will be the only component necessary through the remainder on this work -- note that, for completeness, the other components of $\overleftrightarrow{\chi}_m(\omega)$ are given in the Methods sections. This component is given by 

\begin{equation}
\label{Xxx}
  \chi_{xx}(\omega) = \frac{\chi_a}{1-(\omega/\omega_0)^2}
\end{equation}
where the resonance frequency $\omega_0$ is given by
\begin{equation}
    \omega_0^2 = \gamma^2\mu_0^2[H_0+(D_y-D_z)M_s]\times[H_0+(D_x-D_z)M_s]
\end{equation}
and 
\begin{equation}
\label{Xa}
  \chi_{a} = \frac{M_s}{H_{0}+(D_x-D_z)M_s}.  
\end{equation}

\begin{figure}
\includegraphics[width=0.95\linewidth]{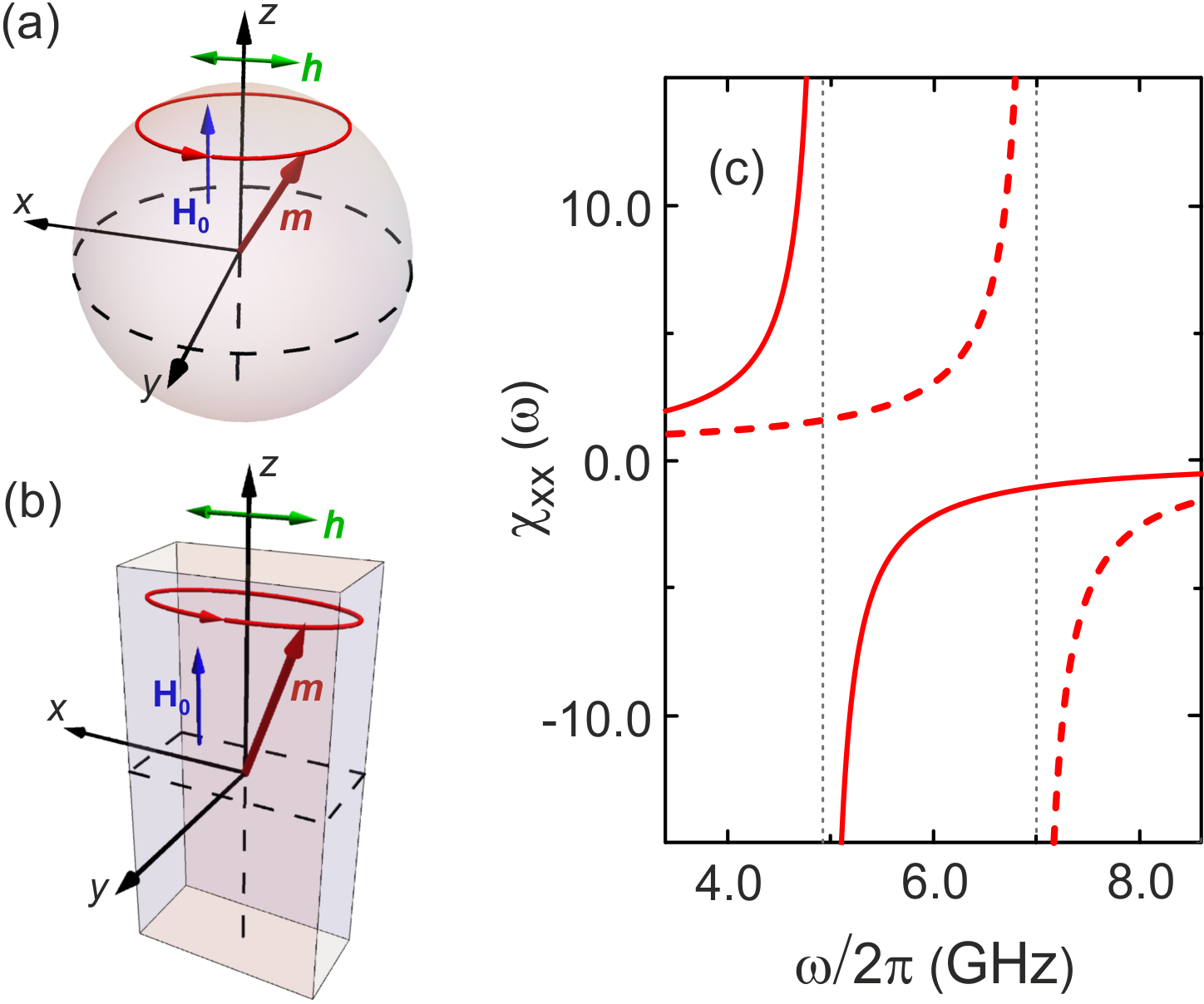}
\caption{
Driving field, $\mathbf{h}$, exciting magnetisation, $\mathbf{m}$, confined to (a) a ferromagnetic sphere and (b) a ferromagnetic rectangular prism. 
(c) Response of $\mathbf{m}$ represented through the susceptibility tensor component $\chi_{xx}(\omega)$. 
The solid lines are for the sphere and the dashed lines are for the rectangular prism. 
The lines for $\chi_{xx}(\omega)$ were calculated using the magnetic parameters for YIG: $\mu_0M_s$~=~0.1758~T, $\gamma/2\pi$~=~28~GHz/T, and $\mu_0H_{0}$~=~0.178~T.
Note that the resonance $\omega_0$ is not the same for the two systems due to different contributions of the demagnetising field.
For a sphere, $\omega_0/2\pi$ = 4.58~GHz as all demagnetising factors are equal to 1/3 due to symmetry \cite{kittel48}.
For a rectangular prism, $\omega_0/2\pi$ = 6.95~GHz and the demagnetising fields were calculated \cite{goppl08} using the dimensions 5$\times$10$\times$50~$\mu$m$^3$ which yield $D_x=0.3266$, $D_y=0.6115$, and $D_z=0.0618$.}
\label{fig:MagPrecess}
\end{figure}

The simplest case to interpret here is that of a ferromagnetic sphere, such as the one depicted in Fig.~\ref{fig:MagPrecess}(a). 
Due to the symmetry of the system, the demagnetising factors are the same in all directions, thus cancelling themselves out in the equations outlined above. 
In this special case, the resonance frequency is now simply $\omega_0 = \gamma\mu_0 H_{0}$, which is the natural precession frequency of a magnetic dipole in a constant magnetic field. 
We can also see that Eq.~(\ref{Xxx}) is reduced to the well know form $\chi_{xx}(\omega) = \omega_m\omega_0/(\omega_0^2-\omega^2)$ with $\omega_m = \gamma\mu_0 M_s$.
For the case of a ferromagnetic rectangular prism [such as the one depicted in Fig.~\ref{fig:MagPrecess}(b)] on the other hand, all demagnetising factors are non-zero \cite{aharoni98} so that both the resonance frequency, $\omega_0$ and permeability tensor components, such as $\chi_{xx}$ have a strong dependence on the components of $\mathbf{H_D}$ as outlined in Eqs.~(\ref{Xxx}-\ref{Xa}).
Note that the demagnetising factors are approximate for rectangular prisms since the demagnetizing fields are in fact nonuniform \cite{aharoni98}.
A comparison between both cases in given in Fig.~\ref{fig:MagPrecess}(c) where the solid lines are for a ferromagnetic sphere and the dashed lines are for a rectangular prism. 
It is then evident that in both case the susceptibility component $\chi_{xx}$ has a singularity at $\omega=\omega_0$.
However, the resonance is shifted to higher frequencies if the demagnetising fields for each direction differ from each other.

It is important to point out that due to symmetry in a sphere $\chi_{xx}(\omega)=\chi_{yy}(\omega)$. 
This is not the case for a rectangular prism with different demagnetising factors along the $x$ and $y$ direction. 
This can be intuitively understood by looking at the cartoons in Fig.~\ref{fig:MagPrecess}(a)-(b). 
For a sphere, the magnetisation is the same in $x$ and $y$, which is in stark contrast to the case of a rectangular prism as $m_x$ and $m_y$ differ from one another. 
Thus, one should expect that $\chi_{xx}(\omega)\neq\chi_{yy}(\omega)$. 
We will not discuss this further as $\chi_{yy}(\omega)$ will not be used in the remainder of this work.
Therefore, we will now move on to the electromagnetic perturbation theory method to describe a cavity-magnon system.

\subsection{Perturbation Theory for Cavity Magnon-Polaritons}
$ $

In practical applications, the resonance frequency of a microwave cavity resonator, $\omega_c$, can be easily modified with the smallest modification in shape, size, or with a small piece of material placed inside the cavity. 
While the effects of these perturbations can often be difficult to quantify, they can be calculated accurately by employing perturbation theory. 
This holds if one assumes that the fields of a cavity with a small shape or material perturbation inside does not greatly deviate from those of the empty cavity.
In recent cavity magnon-polariton experiments, a microwave cavity resonator is modified by introducing a small piece of magnetic material within the cavity.
Up to now, most of the works in cavity spintronics have used approximations or oscillator models to describe the coupling and overall behaviour of the system \cite{harder16,bourhill19}. 
If the magnetic sample is small enough compared to the cavity volume, however, the effects of the sample and the coupling between magnon-cavity can be accurately probed using perturbation theory. 
A short derivation of the most general equations is presented below, and then the results for the specific geometry experimentally studied here are derived. 

We start by looking at an unperturbed cavity state; that of an empty cavity, resonating in only one of its normal modes at frequency $\omega_c$. 
Let the oscillating electric and magnetic fields within the cavity be $\mathbf{E_c}$ and $\mathbf{h_c}$, respectively, proportional to $e^{j\omega_c t}$. 
Under these conditions, one can write Maxwell's equations as:
\begin{subequations}
\label{allE)H)}
 \begin{align}
  \nabla \times \mathbf{h_c} = j\omega_c\varepsilon_0\mathbf{E_c} \label{DxH0} \\
  \nabla \times \mathbf{E_c} = -j\omega_c\mu_0\mathbf{h_c} \label{DxE0}.
 \end{align}
\end{subequations}

On introducing a small ferrite sample into the cavity, the cavity will then resonate at a new frequency $\omega$ \cite{waldron1957}. 
Thus, Eqs. (\ref{DxH0}) and (\ref{DxE0}) have to be rewritten as follows:
\begin{subequations}
\label{allEH}
 \begin{align}
  \nabla \times \mathbf{h} = j\omega\varepsilon_0\mathbf{E} +\mathbf{J_e} \label{DxH} \\
  \nabla \times \mathbf{E} = -j\omega\mu_0\mathbf{h} +\mathbf{J_m} \label{DxE},
 \end{align}
\end{subequations}
where $\mathbf{\mathbf{J_e}}$ and $\mathbf{J_m}$ are the sample's dielectric and magnetic contributions which only exist in the region occupied by the perturbing material and are zero elsewhere in the cavity \cite{fuller1987}.
We can write these quantities as $\mathbf{J_e} = j\omega\varepsilon_0\overleftrightarrow{\chi}_e(\omega)\cdot \mathbf{E} \label{Je}$ and $ \mathbf{J_m} = -j\omega\mu_0\overleftrightarrow{\chi}_m(\omega)\cdot\mathbf{h} \label{Jm}$. 
Here, $\overleftrightarrow{\chi}_e(\omega)$ and $\overleftrightarrow{\chi}_m(\omega)$ are the electric and magnetic susceptibility contributions of the ferrite respectively (both written in tensor form for a more general description).
Following common vector algebraic operations \cite{pozar}, we can obtain the following relation:

\begin{equation}
\label{w-w0Js}
    \omega-\omega_c = j\frac{\displaystyle\int_{\delta v}(\mathbf{J_e}\cdot\mathbf{E_c^*}-\mathbf{J_m}\cdot\mathbf{h_c^*})~\mathrm{d}v}{\displaystyle\int_{v}(\varepsilon_0\mathbf{E_c^*}\cdot\mathbf{E}+\mu_0\mathbf{h_c^*}\cdot\mathbf{h})~\mathrm{d}v}.
\end{equation}
where $\delta v$ is the sample volume, and $v$ is the volume of the empty cavity.

This expression is exact, given the perturbative assumptions made in Eqs.~(\ref{allEH}), and could be evaluated if the configuration of $\mathbf{E}$ and $\mathbf{h}$ for the perturbed cavity were known. 
In general, this can be hard to estimate. 
For cavity measurements in which the samples are small enough, however, one can assume that $\mathbf{E} = \mathbf{E_c}$ and $\mathbf{h} = \mathbf{h_c}$ everywhere inside the cavity. 
For simplicity, and for the remainder of this work, we can also consider that there are no dielectric contributions from the sample and it responds only to the $\mathbf{h_c}$ field of the cavity, so that we can make $\mathbf{E_c} = 0$. This way, Eq. (\ref{w-w0Js}) can be rewritten as 
\begin{equation}
\label{w-w0}
    \omega-\omega_c = -\omega_c\frac{\displaystyle\int_{\delta v}\mu_0\bigg[\overleftrightarrow{\chi}_m(\omega)\cdot\mathbf{h_c}\bigg]\cdot\mathbf{h_c^*}~\mathrm{d}v}{2\displaystyle\int_{v}\mu_0\mathbf{h_c^*}\cdot\mathbf{h_c}~\mathrm{d}v}.
\end{equation}

\section{Dependency on the Distribution of a Linearly Polarised Field}
$ $

\begin{figure*}[ht]
\centering
\includegraphics[width=0.9\linewidth]{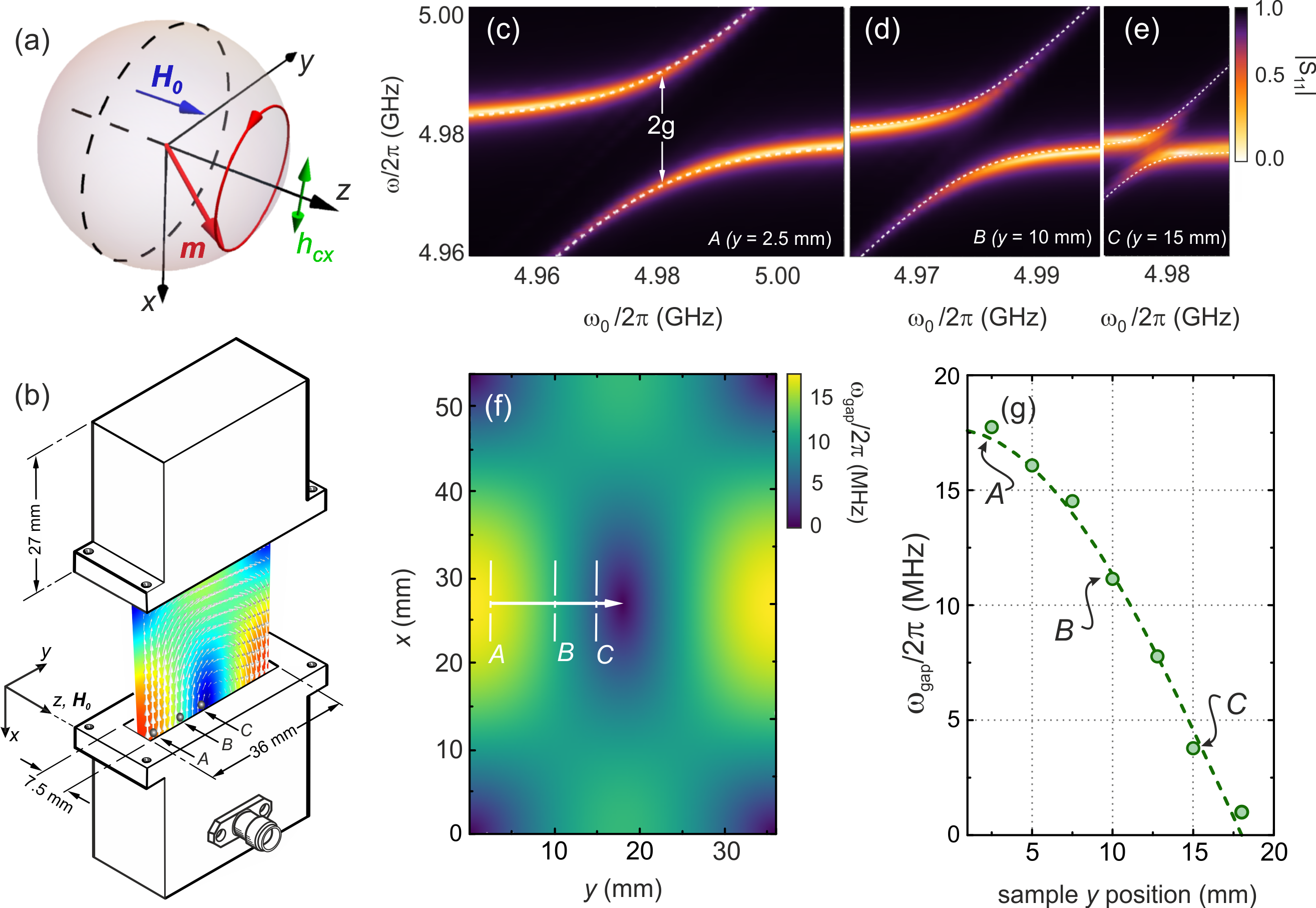}
\caption{(a) Behaviour of magnetisation excited by a linearly polarised excitation such as shown in (b) when the sample is placed inside of a rectangular microwave resonator. 
We show a cross-sectional field configuration at $z$~=~3.75~mm generated by capacitive coupling (simulated with COMSOL). Experimental spectra and perturbation theory (dashed) lines of the Rabi splitting close to $\omega_c$~=~$\omega_0$ a the YIG sphere (0.5 mm diameter) placed at positions (c) $A$ ($y$~=~2.5), (d) $B$ ($y$~=~10), and (e) $C$ ($y$~=~15~mm). 
In part (f) we give a full map of the width of the Rabi splitting, $\omega_{gap}$, for any given $x-y$ position. (g) Experimental points and theoretical lines of $\omega_{gap}$ as the sample is moved within the microwave cavity (along $y$ and at $x$~=~27~mm) through positions $A$, $B$ and $C$ [see panel (b)].}
\label{fig:Field_g_Position}
\end{figure*}

We will now apply the theory detailed so far to understand the coupling between microwaves confined in cavity resonators and magnons. 
We will start by looking at the simple case of microwave fields in a three-dimensional, rectangular cavity exciting magnons in a YIG sphere as depicted in Fig.~\ref{fig:Field_g_Position}(a).

We have experimentally probed the behaviour of the hybrid system as the position of a magnetic sphere inside the cavity is changed. 
Through this, we can gain some insight into how the coupling strength can be modified as we change the profile and configuration of the oscillating field, $\mathbf{h_c}$,  around the sample position. 
For this, we have used a rectangular microwave cavity, such as the one shown in Fig.~\ref{fig:Field_g_Position}(b), with capacitive coupling generating a TE$_{11}$ mode resonating at frequency $\omega_c/2\pi$~=~4.98~GHz. 
The oscillating $\mathbf{h_c}$ intensity profile is also shown in Fig.~\ref{fig:Field_g_Position}(b) with anti-node at $x = 27$~mm, $y = 2.5$~mm and $z = 3.75$~mm, also marked as $A$ (details on the experimental setup are given in Appendix~\ref{app:experiment}).
By placing a small magnetic sample (YIG sphere of diameter 0.5~mm) in the anti-node of $\mathbf{h_c}$ we obtain the Rabi splitting \cite{miller05} displayed in Fig.~\ref{fig:Field_g_Position}(c).
This has often been referred to as level repulsion of the coupled magnon-cavity system and is a classic feature of the hybridisation between these two systems. 
In this case, the macroscopic coupling strength, $g$, is often associated with the width of the splitting at $\omega_c=\omega_0$ which is where the effect of hybridisation is greatest. 
In the strong coupling regime, these are related by $2g = |\omega_a-\omega_b|=\omega_{gap}$ \cite{harder16}. 
Here, $\omega_a$ and $\omega_b$ are the eigenfrequencies for the two modes (branches) seen in Fig.~\ref{fig:Field_g_Position}(c).
The effect of placing the sample away from the anti-node of $\mathbf{h_c}$ is shown in Fig.~\ref{fig:Field_g_Position}(d) for a sample placed at position $B$ ($y = 10$~mm) and in Fig.~\ref{fig:Field_g_Position}(e) for a sample at position $C$ ($y = 15$~mm) where $\omega_{gap}$ is very small as $\mathbf{h_c}$ is close to vanishing -- for reference, the positions A, B, and C are drawn in Fig.~\ref{fig:Field_g_Position}(b).

An efficient way for predicting the behaviour of $\omega_{gap}$ is to employing perturbation theory.  
For this, we consider that in the YIG sphere used in our experiment, the effect of an applied field $\mathbf{H_{0}}$ directed along $z$ is to induce precession that can only couple with the components of $\mathbf{h_c}$ along the $x$ and/or $y$ directions, $h_{cx}$ and $h_{cy}$ respectively.
If we concentrate on the behaviour of the sample moved from the anti-node to the node of $\mathbf{h_c}$ (from $y = 2.5$~mm to point $y = 18$~mm, but always at $x = 27$~mm), we can neglect $h_{cy}$ as it is much smaller than $h_{cx}$ at all points. 
This means the sample is always excited by a linearly polarised field. 
We can then use Eq.~(\ref{mhs}) to rewrite Eq.~(\ref{w-w0}) as:

\begin{align}
\begin{split}
 \label{w-w0usingmu}
    \frac{\omega-\omega_c}{\omega_c} &=-\chi_{xx}(\omega)~\frac{\displaystyle\int_{\delta v}\mu_0|h_{cx}|^2\mathrm{d}v}{2\displaystyle\int_{v}\mu_0\mathbf{h_c^*}\cdot\mathbf{h_c}~\mathrm{d}v}
    \\[2ex]
    {}&=-\frac{\omega_0\omega_m}{\omega_0^2-\omega^2} \frac{W_p}{W_c}.
\end{split}
\end{align}

For simplicity, we write the quantities relating to the oscillating fields in Eq.~(\ref{w-w0usingmu}) as $W_c$ and $W_p$, respectively. 
$W_p$ corresponds to the integral of the oscillating magnetic field across the sample volume and since we can neglect $h_{cy}$ for our positions of interest, in Eq. (\ref{w-w0usingmu}) we were able to reduce this to $W_p = \displaystyle\int_{\delta_v}\mu_0|h_{cx}^2|dv$  -- with $h_{cx}$ being equivalent to $h_x$ in Eq.~(\ref{mhs}).
For a more complicated field geometry involving both x and y components of $\mathbf{h_c}$, see Apendix~\ref{app:Wp}.
As for $W_c$, this represents the total energy stored in the empty cavity ($W_c = W_m+W_e$, where $W_m$ is the total magnetic energy and $W_e$ is the total electric energy). 
In the empty cavity both $W_m$ and $W_e$ are the same, thus we can simply quote one of them, as done in the denominator of Eq.~(\ref{w-w0usingmu}) where we write $W_c = 2 W_m$.

Because we are interested in the behaviour at frequencies close to both the cavity and magnetic resonance frequencies, we can use the relation
$\omega_0^2-\omega^2\approx(\omega_0-\omega)2\omega_0$ into Eq.~(\ref{w-w0usingmu}) to find
\begin{equation}
\label{w-wc_simple}
    (\omega-\omega_c)(\omega-\omega_0) = \frac{1}{2}\omega_c\omega_m\frac{W_p}{W_c}.
\end{equation}
We can then solve this for $\omega$, which yields:
\begin{equation}
\label{w_a_b}
    \omega_{a,b}=\frac{1}{2}\left[\omega_c+\omega_0\pm\sqrt{(\omega_c-\omega_0)^2+2\omega_c\omega_m\frac{W_p}{W_c}}\right].
\end{equation}
These are the eigenfrequencies of the cavity-magnon hybrid system. With these equations, and using the magnetic parameters for YIG, i.e. same as those used in Fig.~\ref{fig:MagPrecess}(c), we obtain the dashed lines in Fig.~\ref{fig:Field_g_Position}(c)-(e) which are in excellent agreement with the experimental contour data.
The intensity of the field at the sample position can be calculated analytic -- with full equations quoted in Appendix~\ref{app:fields}.
For simple rectangular cavities, such as the one shown in in Fig~\ref{fig:Field_g_Position}(a), $W_c$ can also be calculated analytically using $W_c=1/2(\varepsilon_0v)$. 
These relations can be further used to calculate the size of the Rabi splitting, which at $\omega_c=\omega_0$ is given by:
\begin{equation}
\label{DeltaW}
    \omega_{gap}=(\omega_a-\omega_b)\Big|_{\omega_c=\omega_0} = \sqrt{2\omega_c\omega_m \frac{W_p}{W_c}}.
\end{equation}
The heat map shown in Fig.~\ref{fig:Field_g_Position}(f) summarizes the behaviour of $\omega_{gap}$ as function of both $x$ and $y$ positions within the resonator calculated from Eq.~(\ref{DeltaW}). 
This clearly shows that the behaviour of $\omega_{gap}$ strongly reflects the intensity of $\mathbf{h_c}$ given in Fig.~\ref{fig:Field_g_Position}(b).  
Fig.~\ref{fig:Field_g_Position}(g) shows how the predicted values of $\omega_{gap}$ from the analytical expressions from perturbation theory combined with analytical values for the oscillating field inside the cavity (green dashed lines) match experimental points (green dots) as the sample is moved within the resonator shown in part (b) along the $x$-axis.
Note that we have also compared our analytical results with numerical simulations by evaluating $W_c$ and $W_p$ using COMSOL. We have obtained, for position $A$ in Fig.~\ref{fig:Field_g_Position}(b), $W_m=\displaystyle\int_{v}\mu_0\mathbf{h_c^*}\cdot \mathbf{h_c} \mathrm{d}v$~=~5.55$\times$10$^{-10}$~J and $\displaystyle\int_{\delta v}\mu_0|h_{cx}|^2 \mathrm{d}v$~=~6.875$\times$10$^{-15}$~J. 
These values return $W_p/W_c$~=~6.26385$\times$10$^{-6}$
which yield $\omega_{gap}$=17.6~MHz. 
This is consistent with both analytical and experimental values shown in Fig.~\ref{fig:Field_g_Position}(g).

A few main remarks should be made here: 
\begin{itemize}
    \item The coupling constant has been previously estimated by various models fitting experimental data. However, as seen here, perturbation theory is an effective way to exactly calculate $g$ without any need for experimental fitting parameters such as spin density.
    \item In order to solve Eq.~(\ref{w-w0usingmu}), it is not necessary to make any approximations as the ones made here to obtain Eq.~(\ref{w_a_b}). However, the approximations work well close to the splitting.
    \item Finally, at our initial sample position [shown as $A$ in Fig.~\ref{fig:Field_g_Position}(b)] the $\mathbf{h_c}$-field within the cavity is close to its maximum value, and lies in the $\hat{\mathbf{x}}$ direction. 
    While the oscillating field, $\mathbf{h_c}$, gains other components as we move the sample away from the anti-node it is still always linearly polarised; and thus, nothing changes for the perturbation theory. However, the coupling dramatically changes from maximal (at the anti-node of $\mathbf{h_c}$) to vanishing (at the anti-node of $\mathbf{E_c}$). 
\end{itemize}
Furthermore, the perturbation theory described here, with some modifications, can be readily applied to microwave resonators and transmission lines of any kind by describing the field distribution. It can also be used for a variety of different magnetic samples by obtaining the appropriate permeability tensors (considering shapes, structuring, or composition). A case example is now given.

\section{Magnetic Thin-Film Stripes Coupled to a Transmission Line Resonator}
$ $

\begin{figure*}
\centering
\includegraphics[width=0.95\linewidth]{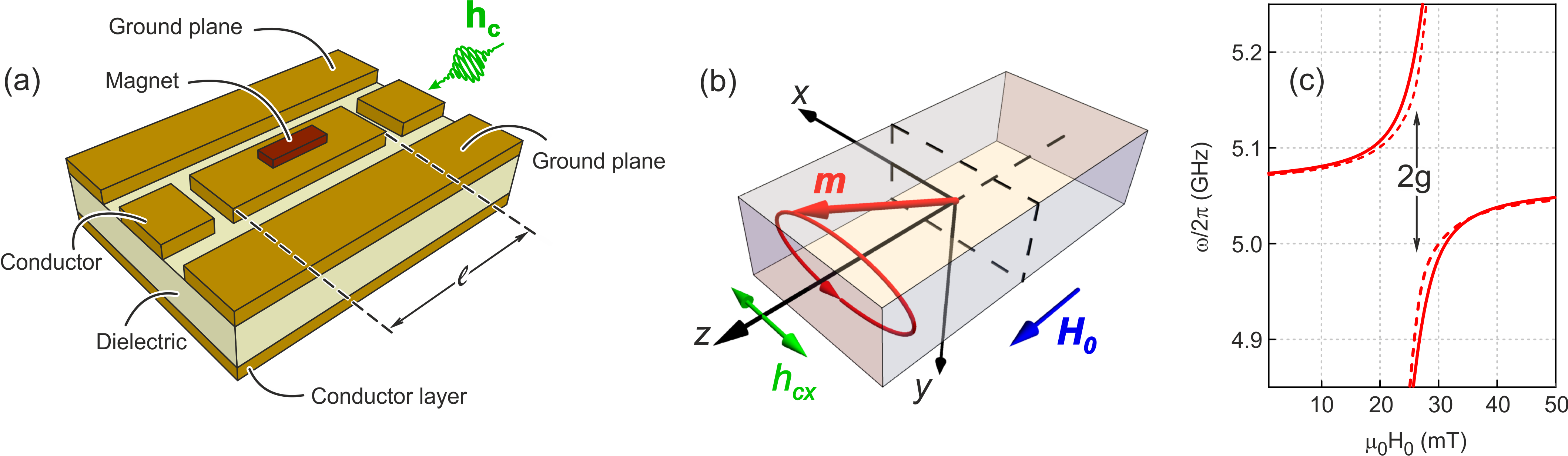}
\caption{(a) Diagram of the set up used here where a two-dimensional coplanar waveguide resonator generates an oscillating magnetic field which couples to oscillating magnetisation in a magnetic thin-film stripe. 
A static field, $\mathbf{H_{0}}$ is applied along the sample's long axis (along $z$) and the oscillating magnetic field, $\mathbf{h_c}$, at the sample position only has a component along the $x$ direction -- denoted as $h_{cx}$. 
A full schematic of the magnetic sample and oscillating magnetisation with respect to the oscillating and static fields is given in (b). 
(c) Spectra of the hybrid magnon-resonator modes calculated close to the Rabi splitting ($\omega_c$~=~$\omega_0$).
Here, we considered the magnetic sample to be a Py (Ni$_{80}$Fe$_{20}$) rectangular prism (14$\times$0.03$\times$900~$\mu$m$^3$) and the resonance frequency of the resonator is $\omega_c/2\pi$~=~5.0~GHz. The solid lines are for no damping [using Eq.~(\ref{w_a_b_demag})] and the dashed lines take damping into account.}
\label{fig:2Dresonator}
\end{figure*}

Now that we have investigated a simple case, against which we have verified the validity of the perturbation theory for various field configurations, we move on to a more complicated resonator.
Up to now we have discussed the case of a large and three-dimensional system, where the resonator is in the order of a few centimetres. 
This is often the dimensions of cavity spintronic devices to enhance the coupling rates. 
However, in order to easily integrated these with either silicon-based or superconducting quantum circuits, for example, it is necessary to reduce the system's dimensions towards on-chip scalable devices.
The first studies on new on-chip cavity spintronic devices were by Hou et al. \cite{hou2019} and Li et al. \cite{li19}.  
We will therefore now look at a similar system to the ones investigated in those studies, where a micrometre-sized stripe of magnetic materials is placed on planar (two-dimensional) superconducting resonators.
One way to get this is by employing a coplanar waveguide structure with two gaps in the center conductor such as the schematics shown in Fig.~\ref{fig:2Dresonator}(a). 
The gaps in the centre conductor form a planar capacitor acting as dielectric mirrors, which in turns generates standing waves. These determine the resonance frequency through the separation length between the capacitors, $l$, which is a multiple of the half wavelength $\lambda$/2 \cite{goppl08}.

The magnetic thin-film stripe is placed in the center of the resonator, and much like our previous case of a three-dimension system, an external magnetic field is used to set the ferromagnetic frequency of the sample near the cavity resonance frequency.
However, as opposed to the case of a sphere, the magnetic precession drastically changes due to the shape of the sample, as shown in Fig.~\ref{fig:2Dresonator}(b) and discussed in Sec.~\ref{sec:LLG}.
The confined dimensions now induce highly elliptical precession behaviour and that can be quantified through demagnetising factors in the susceptibility, as discussed in Eq.~(\ref{mhs}).
In order to obtain the condition for ferromagnetic resonance, the sample is positioned so that the oscillating magnetic field generated by the centre conductor is perpendicular to the static field. 
Here, the relevant component of the oscillating field $\mathbf{h_c}$ at the sample position is, again, along the $x$ direction [in Fig.~\ref{fig:2Dresonator}(b) this is depicted as $h_{cx}$].  

The more complex field profiles inside the two-dimensional resonator, compared to the 3D cavity discussed earlier, only marginally affect the perturbation method described earlier. 
In fact, the main difference for this particular case is the calculation of the fields exciting the magnetic sample, contained in $W_p$, and the total energy stored in the resonator $W_c$ [both discussed near Eq.~(\ref{w_a_b})].
These quantities can no longer be calculated analytically as we have done in previous sections, but they can easily be estimated using electromagnetic solvers such as HFSS or COMSOL (See supplemental information for details). 
Once those are estimated, the eigenfrequencies can be obtained using:

\begin{equation}
\label{w_a_b_demag}
    \omega_{a,b}=\frac{1}{2}\left[\omega_c+\omega_0\pm\sqrt{(\omega_c-\omega_0)^2+2\chi_a\omega_c\omega_0 \frac{W_p}{W_c}}\right].
\end{equation}

\noindent Note that this equation is slightly different from Eq.~(\ref{w_a_b}). 
This is because we now have to account for the demagnetising fields and use the full form of $\chi_{xx}$ as given in Eq.~(\ref{Xxx}).
The resulting Rabi splitting calculated using Eq.~(\ref{w_a_b_demag}) is shown in Fig.~\ref{fig:2Dresonator}(c) as the solid lines. 
For this, we have used Ni$_{80}$Fe$_{20}$ -- also known as Permalloy (Py) -- as the example material for the magnetic thin-film stripe with parameters $\mu_0M_s=1$~T and $\gamma/2\pi = 28$~GHz/T \cite{li19}.
The demagnetising parameters are $D_x$~=~0.0052, $D_y$~=~0.9947, and $D_z$~=~0.00008 \cite{aharoni98} and the oscillating magnetic field $h_{0x}$ at the sample position and energy stored in the system $W_c$ were calculated using COMSOL yielding $W_p$~=~2.868$\times$10$^{-15}$~J and $W_c$~=~3.6$\times$10$^{-11}$~J for a ration $W_p/W_c$~=~7.967$\times$10$^{-5}$.

Knowing the fields in the resonator and the dimensions of the magnetic sample, it is also straightforward to estimate the coupling constant $g$ through the width of the splitting using the relation:
\begin{equation}
\label{DeltaWdemag}
    \omega_{gap}=(\omega_a-\omega_b)\Big|_{\omega_c=\omega_0} = \sqrt{2\chi_a\omega_c\omega_0 \frac{W_p}{W_c}}.
\end{equation}
This yields $\omega_{gap}/2\pi =$~0.357~GHz or $g/2\pi=$~0.178 GHz. 

\subsection{Effect of Damping on the Coupling Strength}
$ $

Our calculated coupling strength, $g$, using Eq.~(\ref{DeltaWdemag}) is 26.6~MHz higher when compared to the case reported by Li and co-workers \cite{li19}.
While we have used the same resonance frequency, material parameters, and sample dimensions as reported in their work, there is one property we have neglected so far: damping.
This was not necessary when looking at YIG spheres, as in this case the linewidth of the magnetic resonance is small enough so that it does not affect the eighenfrequencies from perturbation theory. 
In magnetic thin films, however, such linewidths are not only a result of intrisinc damping but they are often broadened by various surface and interface non-uniformity as well as sample defects -- known as inhomogeneous broadening. 
In general, the effect of damping and dissipation can be introduced by replacing $\omega_0$ with a complex frequency $\omega_0\rightarrow\omega_0'+j\omega_0''$, or even by just replacing $H_{0}$ with a complex magnetic field $H_{0}\rightarrow H_{0}'+j\Delta H_{0}$ where $\Delta H_0$ is the width of the resonance curve at half height. 

Applying the former description into Eq.~(\ref{w-w0usingmu}) we obtain the dashed lines in Fig.~\ref{fig:2Dresonator}(c) using $\omega_0''=$~0.035~GHz as measured by Li and co-workers \cite{li19}. 
The Rabi splitting for a Py thin-film strip when dissipation is considered is clearly smaller than when no damping is taken into account.   
With dissipation, Eq.~(\ref{w_a_b_demag}) which quantifies $\omega_{gap}$ now becomes:
\begin{multline}
\label{Dw_demag_damp}
     \omega_{gap}=(\omega_a-\omega_b)\Big|_{\omega_c=\omega_0} =
\\[2ex]
         \sqrt{2\chi_a\omega_c\omega_0' \frac{W_p}{W_c}+j\omega_0''\left(2\chi_a\omega_c\frac{W_p}{W_c}+j\omega_0''\right)},
\end{multline}

\noindent and for the case shown in Fig.~\ref{fig:2Dresonator}(c), we can use Eq.~(\ref{Dw_demag_damp}) to find $\omega_{gap}/2\pi$~=~0.309~GHz or simply $g/2\pi$~=~0.154~GHz. 
This is in excellent agreement with the value for the coupling strength of $g/2\pi$~=~0.152~GHz reported by Li and co-workers \cite{li19}.

\section{Scattering Parameter and Quality Factor from Perturbation Theory}
$ $

As we have seen so far, cavity perturbation theory in itself is an extremely efficient method to measure the coupling strength of hybrid systems. 
However, this technique can also be employed when computing scattering parameters.
These quantities are often measured by VNAs in spectroscopic experiments, much like the data we have discussed in Fig.~\ref{fig:Field_g_Position}(c)-(e).

 \begin{figure*}
\centering
\includegraphics[width=0.7\linewidth]{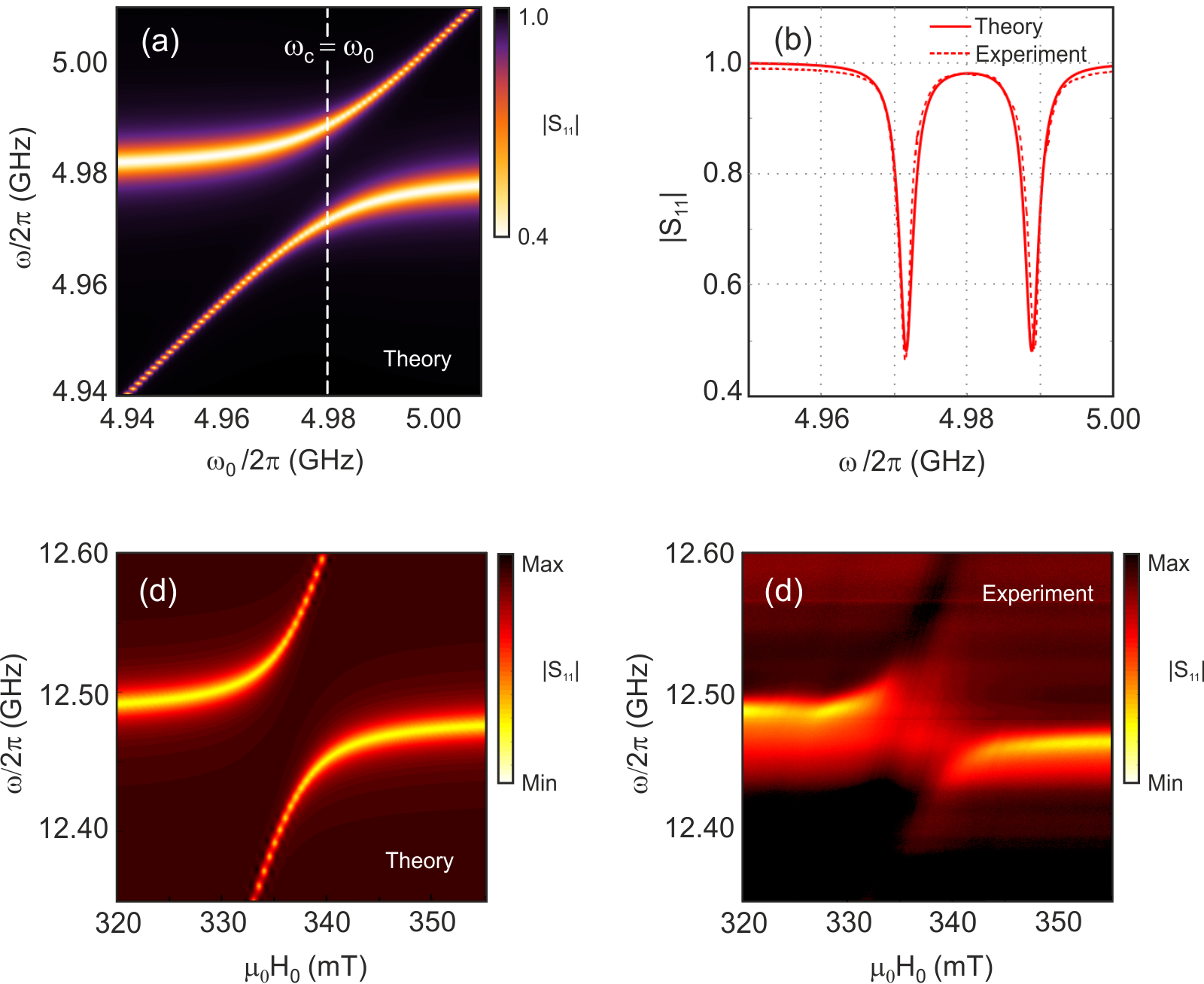}
\caption{(a) Scattering parameter $|S_{11}|$ calculated from the quality factor, $Q_p$, from perturbation theory as a function of both input frequency $\omega$ and externally applied magnetic field $\mathbf{H_{0}}$. (b) Comparison between experimental and theoretical $|S_{11}|$ spectra at $\omega_c$~=~$\omega_0$. The solid line is for the theory [vertical cut in (a)] and the dashed line is for experimental data [a vertical cut in Fig.~\ref{fig:Field_g_Position}(b)]. 
Here we have considered the system to be slightly overcoupled with $\beta$~=~$1.05$; and the dissipation for the two systems were taken to be $\omega_c''$~=~$10^{-3}$ and $\omega_0''$~=~$10^{-4}$.
Scattering parameters for a two-dimensional microwave resonator coupled to a YIG thin-film stripe are also given using perturbation theory (c) compared to experimental data (d).
}
\label{fig:S11fromQ}
\end{figure*}

We can then employ a scattering matrix formalism in order to investigate how microwave radiation interacts with
the hybrid system.
In the vicinity of the resonances, the behavior of
the cavity resonator and magnet can both be represented as lumped circuits.
This way we can assume that a voltage wave $a_1$ is incident on an arbitrary microwave device. 
The wave is scattered and some of its energy goes into a reflected wave $a_2$ and part into a transmitted wave $b_2$. 
Therefore the scattering parameter $S_{11}(\omega)$ (which we have looked at Fig.~\ref{fig:Field_g_Position}) is given by the ratio $a_2/a_1$ \cite{han96}. 
Moreover, in order to account for both resonant hybrid systems, we take into account the product of the two resonances obtained as our perturbation theory eigenfrequencies so that we can make $S_{11}(\omega) = S^{(a)}_{11}(\omega)\times S^{(b)}_{11}(\omega)$, with \cite{luiten05}:

\begin{equation}
    S^{(a,b)}_{11}(\omega) = \frac{\beta-1-jQ_{(a,b)}[\omega/\omega_{(a,b)}-\omega_{(a,b)}/\omega]}{\beta+1+jQ_{(a,b)}[\omega/\omega_{(a,b)}-\omega_{(a,b)}/\omega]}.
\end{equation}

\noindent Here, $\beta$ is the propagation
constant which determines whether the system is \textit{undercoupled} (taking $\beta <1$); \textit{overcoupled} (taking $\beta >1$); or if the resonator is critically coupled (taking $\beta =1$). We also take the quality factor $Q$ to be \cite{probst15}: 

 \begin{equation}
 Q_{(a,b)}=\frac{\omega_{(a,b)}'}{\omega_{(a,b)}''},    
 \end{equation}
 
 \noindent where $\omega_{(a,b)}'$ is the real part of either eigenfrequency $a$ or $b$ calculating with the equations from perturbation theory, such as Eq.~(\ref{w_a_b}), and $\omega_{(a,b)}''$ denotes their equivalent imaginary part. 
 
In Fig.~\ref{fig:S11fromQ}(a) we show the calculated scattering parameter mirroring that shown in Fig.~\ref{fig:Field_g_Position}(c) at the node of the magnetic field inside the resonator.
In Fig.~\ref{fig:S11fromQ}(b) a comparison between the experimental and theoretical $|S_{11}|$ parameters obtained using perturbation theory at $\omega_c=\omega_0$ is given.
Note that to obtain those it was necessary to consider dissipation for both cavity and magnetic systems - much like the case we have just discussed for a thin-film stripe. 
This can again be accounted for by making $\omega_c = \omega_c'+j\omega_c''$ and $\omega_0 = \omega_0'+j\omega_0''$, where $\omega_c''$ and $\omega_0''$ are the dissipation in the cavity and in the magnetic sample, respectively. 

The same principles outlined above can also be applied to more complex resonator systems, as is the case of two-dimensional resonators. 
In Fig.~\ref{fig:S11fromQ}(c) we show the theoretical spectra for a thin-film YIG stripe coupled to a Niobium Nitride (NbN) superconducting in-plane resonator. 
The YIG sample size is 5$\times$10$\times$50~$\mu$m$^3$ with magnetic properties as discussed in Fig.~\ref{fig:MagPrecess}. The eigenfrequencies were obtained as done in the previous section -- using Eq.~\ref{w_a_b_demag}.
For comparison, in Fig.~\ref{fig:S11fromQ}(d) we show the experimental spectra for the same set up.
Here, the YIG sample was integrated into the planar superconducting devices using a Xenon Plasma Focused Ion Beam (PFIB) and the spectra was measured at 3.2~K -- the full sample preparation and measurement is detailed  in Ref. \cite{baity20}.
This is in excellent agreement with our theoretical data shown in Fig.~\ref{fig:S11fromQ}(c).

\section{Conclusion}
$ $

To fully understand the electromagnetic behaviour of cavity-magnon hybrid systems we have employed a versatile and self-consistent theory which is an excellent tool to estimate the coupling strength $g$ (or width of the Rabi splitting $\omega_{gap}$) without fitting parameters such as spin density.
This technique allows us to describe the direct interaction between the microwave excitation vector fields and magnetisation dynamics.
The direction, profile, and intensity of microwave fields can dramatically alter magnon-cavity hybridised states, i.e. cavity magnon-polaritons. 
The understanding presented here is particularly relevant for technological applications based on cavity spintronics.
For instance, these systems are expected to aid bidirectional conversion between radio-frequency waves and light \cite{hisatomi16,lambert19}. 
Moreover, as cavity magnon–polaritons can also couple with qubits \cite{Lachance-Quirion20}, they are also expected to be used as an aid to quantum information processing \cite{tabuchi15,lachance_quirion19}. 
In both cases, engineering as well as understanding the coupling are crucial steps to optimise the conversion and (or) information exchange. 

In our work, through perturbation theory, we are able to predict as well as gain further insight into the nature of the coupling between microwave cavities and magnons in a rigorous manner for any resonator with any field configuration and for any geometry of the sample (e.g. spheres or thin films, pillars, etc). 
Our theory is particularly relevant, not only from a fundamental point of view, but also practically, as in order to engineer and optimise cavity spintronic devices the behaviour of the coupling must be fully understood. 
While the constant $g$ is more often obtained from experimental data and incorporated into models, such as the circuit model \cite{harder16}, there have been efforts to describe the coupling with no fitting parameters \cite{bourhill19,zhang14}.
This however, is done through phenomenological oscillator models where the total spin number is key. 
The work presented here employs the magnetic susceptibility of a sphere and rectangular prism to find the magnon-cavity coupling.
The magnetic susceptibility can be found for vastly more complicated magnetic systems, where exchange interactions and dipolar interactions are important to consider. 
This involves solving the Landau-Lifshitz equation numerically for an array of spins, rather than a macrospin as was done here. 
Accurate prediction is therefore possible for the coupling strength between electromagnetic cavity waves and magnons in samples with complicated shapes or spin orderings.
Our findings show excellent agreement with recently published works towards miniaturisation of hybrid systems and provide a new avenue to predict the coupling to not only extremely low damping magnetic films such as YIG \cite{baity20}, but also to highly damped metallic thin films \cite{li19,hou2019}.  
\newline

\begin{acknowledgments}
We would like to thank Robert E. Camley, Tim Wolz and Isabella Boventer for their most useful discussions as well as their comments on the manuscript.  
This work was supported by the European Research Council (ERC) under the Grant Agreement 648011, the Initiative and Networking Fund of the Helmholtz Association, the Leverhulme Trust and the University of Glasgow through LKAS funds. 
D. A. Bozhko acknowledges support from the Alexander von Humboldt Foundation.
R. Holland was supported by the Engineering and Physical Sciences Research Council (EPSRC) through the Vacation Internships Scheme.
\end{acknowledgments}

\appendix

\section{Full Susceptibility Tensor}
$ $

Further to $\chi_{xx}$ -- which was given in the main text as Eq.~(\ref{Xxx}) -- the other components of the susceptibility tensor, $\overleftrightarrow{\chi}_m(\omega)$, given in Eq.~\ref{mhs} are:
\begin{subequations}
\label{XyyNXxy}
 \begin{align}
  \label{Xyy}\chi_{yy}(\omega) = \frac{\chi_b}{1-(\omega/\omega_0)^2}\\
  \label{Xxy}\chi_{xy}(\omega) = \frac{\omega\omega_m}{\omega_0^2[1-(\omega/\omega_0)^2]}.
 \end{align}
\end{subequations}
where 
\begin{equation}
    \chi_{b} = \frac{M_s}{H_{0}+(D_y-D_z)M_s}.
\end{equation}

\section{Fields in a Rectangular Cavity Resonator}\label{app:fields}
$ $

The rectangular cavity resonator shown in Fig.~\ref{fig:Field_g_Position}(b) includes height ($a$~=~54~mm), length ($b$~=~36~mm), and width ($c$~=~7.5~mm). We have concentrated on a TE$_{mn}$ propagation mode where the subscripts $m$ and $n$ come from the wave propagation modes and represent the changing cycles along $x$ and $y$. 

On the basis of our discussion on a TE propagation mode of a rectangular cavity and the boundary conditions, the field distribution of these modes can be written as:

\begin{subequations}
\label{suplemental fields}
 \begin{align}
h_{cx} =j \frac{\kappa_{0y}}{\omega_c\mu_0} \sin \kappa_{0x}x\cos\kappa_{0y}y  \\
h_{cy} =-j \frac{\kappa_{0x}}{\omega_c\mu_0} \cos \kappa_{0x}x\sin\kappa_{0y}y,
 \end{align}
\end{subequations}
with
\begin{equation*}
\kappa_{0x}=\frac{m\pi}{a}, \kappa_{0y}=\frac{n\pi}{b},
\kappa_{0x}^2+\kappa_{0y}^2 = \kappa_{0}^2 = \omega^2\varepsilon_0\mu_0.
\end{equation*}

Note that these field profiles can also be calculated using electromagnetic solver such as COMSOL or HFSS. 

\section{General form of $W_p$}\label{app:Wp}

In the most general case, where the excitation vector fields lie in both $x$ and $y$ directions, the numerator of Eq.~(\ref{w-w0}) is now given by:

\begin{multline}
\label{Wp_general}
    \displaystyle\int_{\delta v}\mu_0\Big[\chi_{xx}(\omega)|h_{cx}|^2+\chi_{yy}(\omega)|h_{cy}|^2\\
    +j\chi_{xy}(\omega)((h_{cy}h_{cx}^*-h_{cx}h_{cy}^*)\Big]dv,
\end{multline}

or simply:

\begin{multline}
       \displaystyle\int_{\delta v}\mu_0\frac{\omega_0\omega_m}{\omega_0^2-\omega^2} \Bigg[  |h_{cx}|^2+|h_{cy}|^2\\
    +j\frac{\omega}{\omega_0}((h_{cy}h_{cx}^*-h_{cx}h_{cy}^*) \Bigg] dv.   
\end{multline}

Thus, in the limit close to $\omega=\omega_0$, we can write $W_p$ as simply:
\begin{equation}
    W_p = \displaystyle\int_{\delta v}\Big[|h_{cx}|^2+|h_{cy}|^2+j((h_{cy}h_{cx}^*-h_{cx}h_{cy}^*)\Big]dv,
\end{equation}
making the current form of Eqs.(\ref{w-wc_simple})-(\ref{DeltaW}) does not change, apart from the value of $W_p$ to incorporate the different direction of $\mathbf{h_c}$.
Note that these equations hold true for a sphere only. They would need to be modified with the relevant susceptibility tensor components for other shaped systems considering the effects of demagnetising fields. For instance, for a rectangular prism, as that shown in Fig.~\ref{fig:MagPrecess}(b), one can simply take Eqs.~(\ref{Xxx}),(\ref{Xyy}), and (\ref{Xxy}) and plug them into Eq.~(\ref{Wp_general}) to find the appropriate numerator of Eq.~(\ref{w-w0}). 

\section{Experimental Setup Details}\label{app:experiment}
$ $

For the three-dimensional rectangular resonator measurements, microwave signals were supplied by port 1 of a Rohde \& Schwarz ZVA 40 vector network analyser (VNA) and signals reflected from or transmitted through the cavity were sent to port 2 of the VNA.
The capacitive coupling to the cavity was tuned by adjusting the length of the SMA connector contacts, which extended into the body of the cavity.

%

\end{document}